\documentclass{article}
\usepackage[dvips]{graphicx}
\def\be{\begin{equation}}
\def\ee{\end{equation}}
\def\bea{\begin{eqnarray}}
\def\eea{\end{eqnarray}}
\def\d{{\rm d}}

         \def\d{{\rm d}}

         \def\l{\lambda}
         
         \def\la{\label}
         \def\rf{\ref}

         \def\se{\section}
         \def\sse{\subsection}
         \def\o{\over}
         \def\a{\alpha}
         \def\b{\beta}
         \def\t{\tau}

\begin{document}
\begin{titlepage}
\vspace{1cm}
\begin{center} {\Large \bf  Exact determination of the phase structure
of a \\
multi-species asymmetric exclusion
process}\\
\centerline {\bf ${\rm M. Khorrami}^{a,c}$ \footnote {e-mail:mamwad@theory.ipm.ac.ir}
\hskip 1.5cm ${\rm V. Karimipour}^{b,c}$ \footnote{e-mail:vahid@theory.ipm.ac.ir}}
{\it $ ^a$Institue for advanced studies in basic sciences, Gava Zang},
P.O.Box 45195-159
, Zanjan, Iran\\
{\it $ ^b$Department of Physics, Sharif University of Technology},
P.O.Box 11365-9161, Tehran, Iran \\
{\it $^c$Institute for Studies in Theoretical Physics and Mathematics},
P.O.Box 19395-5746, Tehran, Iran.\\
\end{center}
\vskip 2cm
\begin{abstract}
\noindent
We consider a multi-species generalization of the Asymmetric Simple Exclusion
Process on an open chain, in which particles hop with their
characteristic
hopping rates and fast particles can overtake slow ones.
The number of species is arbitrary and the hopping rates can be selected from
a discrete or continuous distribution. 
 We determine exactly
the phase structure of this model and show how the phase diagram of the
1-species ASEP is modified. Depending on the distribution of hopping rates, the system
can exist in a three-phase regime or a two-phase regime. In the three-
phase regime the phase structure is almost the same as in the one species
case, that is, there are the low density, the high density and the maximal current
phases, while in the two-phase regime there is no high-density phase.
\end{abstract}
\end{titlepage}
\section{Introduction}
The asymmetric simple exclusion process \cite{sz,sp,l,sph,dh,d} refers to a collection of Brownian particles which under the influence of a
driving force, do biased random hopping
on a one dimensional lattice and interact via hard core repulsion with each
other. In the totally asymmetric case each particle is injected to the system
from the left with rate $\a$ and hops only to the right neighboring site with
a rate normalized to unity and finally is extracted at the right end with rate
$\b$.\\
This is a model far from equilibrium with many re-interpretations which makes it
a suitable model for studying such phenomena as diverse as surface
growth \cite{kr}, and traffic flow \cite{hh} \cite{hs},
(see \cite{er},\cite{deg}
,\cite{d,de} and references
therein.)\\
One of the most interesting aspects of this process is the possibility of
boundary induced phase transitions. It has been observed through various
types of solutions \cite{ligg, krug, sch,schd,dehp} that by changing
the rate of injection and extraction of particles, different phases
will develop
in the system. The phase diagram of the model representing the macroscopic current
of the particles in various domains in the $\a-\b$ plane is depicted
in fig.(1).\\
\begin{figure}
\centering
\includegraphics[height=10cm,angle=00]{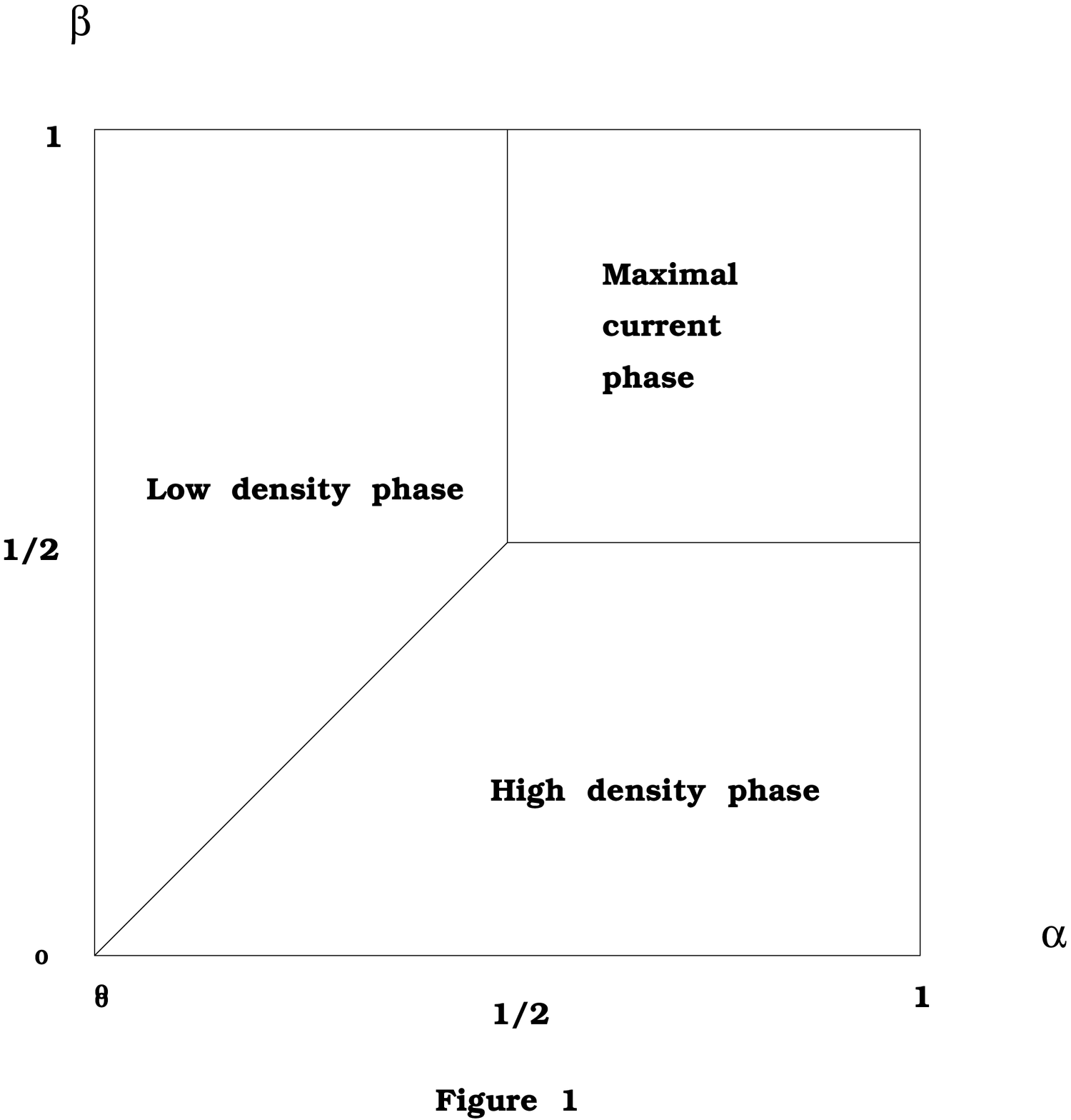}
\caption{Phase diagram for the single species ASEP}
\end{figure}

Although some of the characteristics of the process have been obtained 
by other analytical methods,\cite{sch}
\cite{schd},\cite{ddm}, the solution by Matrix Product Ansatz (MPA) \cite{dehp}
has proved much useful for obtaining among other things, the same results in
much simpler ways.
The general formulation of MPA \cite{ks} has also been shown to be amenable 
to further generalizations\cite{ahr,adr,k1,k2,ev,evg,lpk,mmr,fj}.\\
In the MPA the probabilities $P(\t_1\cdots\t_L)$ , where $\t_i$ is the random
variable associated with site $i $ (being $0$ for empty site and $1$ for an
occupied
site), is written as:
\be \la{v1}P(\t_1,\cdots\t_L)= {1\over Z_L} <W|\prod_{i=1}^{L}(\t_i D +
(1-\t_i)E|V>\ee
where the operators $ D $ and $ E$, and the vectors $ <W|$ and $|V>$ satisfy
the following relations:
\be \la{v} DE = D + E \ee
\be \la{vv} D|V> = {1\over \b}|V>\ee
\be \la{vvv}<W|E = {1\over \a}<W|\ee
In (\rf {v1}), $ Z_L $ is a normalization constant and is suitably called the
partition function. Its value is given by
\be \la{v3} Z_L = <W| C^L|V> \ee
where $ C:=D+E$.\\
The question of a natural $p-$ species generalization of the ASEP, so that
in the special case $ p=1 $, one obtains the results of the one species ASEP
\cite{adr}, has been answered in the affirmative in \cite {k1,k2}, by
postulating a generalization of the algebra (\rf{v}-\rf{vvv}),
which we call the
$p-$ASEP algebra.
Since the number of species is quite arbitrary and it can even be infinite, in
which case the hopping rates are taken from a continuous distribution, we rewrite
the multi-species algebra in a more general form than that of \cite{k1}.
This new algebra is generated by a discrete generator $E$ and a one parameter
family of generators $ D(v)$ where $ v \in R^+$. These generators are subject
to the following relations:
\begin{eqnarray}\la{v4} D(v) E &=& {1\over v} D(v) + E \\ \la{v4'}
D(v') D(v) &=& {v D(v') - v'D(v) \over{v - v'}} \ \ \ \ \ \  v'> v \\ \la{v4''}D(v) |V>&=& {v\over {v+\b-1}} |V> \\
\la{v4'''}<W|E&=& {1\over \a}<W|\end{eqnarray}
We also need to show how this continuously parameterized algebra is derived
in the MPA formalism. This is done in the appendix.
The hopping rates $v$ are taken from a general distribution $\sigma(v)$,
with support $ [v_1,\infty)$. That is, $v_1$ is the smallest hopping rate
in the ensemble. Note that this ensemble refers to the particles waiting
to enter the system, or the ensemble of particles moving in the system, if
there were no interactions between the particles. For this reason we call
$v$ the intrinsic hopping rate or average velocity of a particle.\\
The process described by this algebra is one in which each particle of velocity
$v$ arrives at the left end with rate $ {\a}(v):= \a v $, (i.e: the input current
is $ \sigma(v) \a v$),
hops to its right
neighboring empty site with rate $ v $ and leaves the system at the right
end with rate $ {\b}(v) =\b+v - 1 $. If this particle encounters on its way
a site occupied by a particle of intrinsic velocity $v'$ with $v'< v$, it will overtake it
with rate $v - v' $, otherwise it stops. For all the extraction rates $\b(v)$
to be positive we
also require that $ \b\geq 1-v_1$. The unit of time is set so that the average
hopping rate is unity, i.e: $ \int v \sigma(v)dv = 1 $.
Thus the parameters $\a$ and $\b$ are respectively the total injection rate and
the average extraction rate of the particles respectively.
Note that although the multiplicative dependence of the injection rate
$\a (v)$ on $ v$ is rather natural, this is not the case for the extraction
rate $ \b(v)$ the form of which is dictated only by our demand to solve the system through
the MPA relations (\rf{v4}-\rf{v4'''}).
Note also that all the elementary processes are stochastic, i.e: in a time
interval $dt$, a particle of velocity $v$ present in a given site, hops to the
right empty site
with probability $ v dt$, and does not move with probability $ 1-v dt$.\\
The model we consider depends on two boundary parameters $\a$ and $\b$
and on the distribution function
$ \sigma(v)$.
All our arguments below
are also valid for a discrete distribution, for which $ \sigma(v):={1\o p}\sum_{i}
\delta(v-v_i)$.\\

The main motivation for pursuing this problem is to see how the phase structure of
the one-species ASEP (hereafter denoted by 1-ASEP) is modified, when we have particles with a variety of
hopping rates and especially when particles can overtake each other.
Do we still have the  phases of low-density, high-density and maximum
Currents, present in the 1-ASEP, or is it changed in an essential way?.
How does the variety of
hopping rates in the bulk or their probability distribution enter the
picture
and what role does this distribution play in the phase structure of the system ? How the absence of particle-hole symmetry in this model is reflected in the phase diagram?\\
We will go through
these questions by providing an exact solution of this problem, and will obtain
a generalization of the phase diagram of the 1-ASEP.\\ 
As far as we consider only the mean field line $\a + \b = 1$
\cite{k1}, one dimensional representations of the algebra (\rf {v4}-\rf{v4'''})
give
an exact solution. However to uncover the important role of fluctuations,
we should explore the full $ \a-\b$ plane and for this we should use
the infinite dimensional representation. What we will do is to calculate
exactly the generating function for partition functions of systems with different
sizes and by carrying out an analysis of its singularities, determine the
currents and the different phases of the system.\\
The phase structure depends on the values of $\a$, $\b$, and on the characteristics
of the distribution function.\\

{\bf \large {Main Results:}}\\
${\bf \bullet }$ 
To every distribution function $\sigma (v)$ of hopping rates, we can assign a real
number $l[\sigma]$, defined in (\rf{dis}), 
which essentially depends on the behavior of $ \sigma(v)$ for
small hopping rates (i.e: if $\sigma (v_1)=0$ or not and if yes how slowly it approaches
this value). The parameter $l_c = 0$ is special in the sense that for all
distribution functions with $l[\sigma]<0$, the phase
diagram of the multi-species ASEP is almost the same as the phase diagram of 1-ASEP,
that is, in the $\a-\b$ plane we have three regions of low density, high density
and maximum current phases. The value of the maximum current and the shape
of the coexistence curves between different phases depend
on the distribution function (see figs.(2)and (3)). We also obtain the average density
of all types of particles in all three phases.\\

\begin{figure}
\centering
\includegraphics[height=10cm,angle=00]{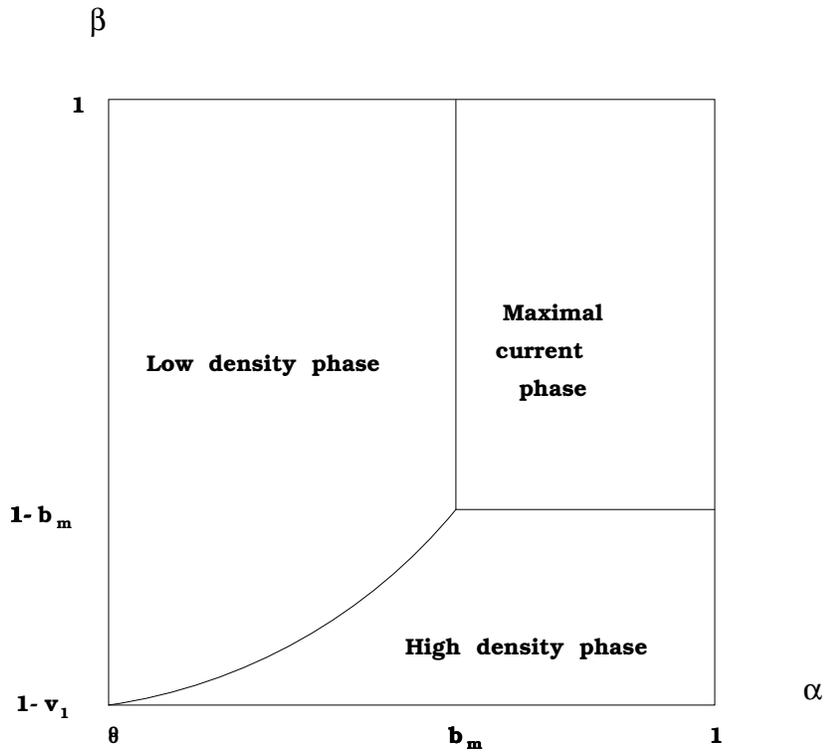}
\caption{Phase diagram for multi-species ASEP when $l[\sigma]<0$}
\end{figure}
\begin{figure}
\centering
\includegraphics[height=10cm,angle=00]{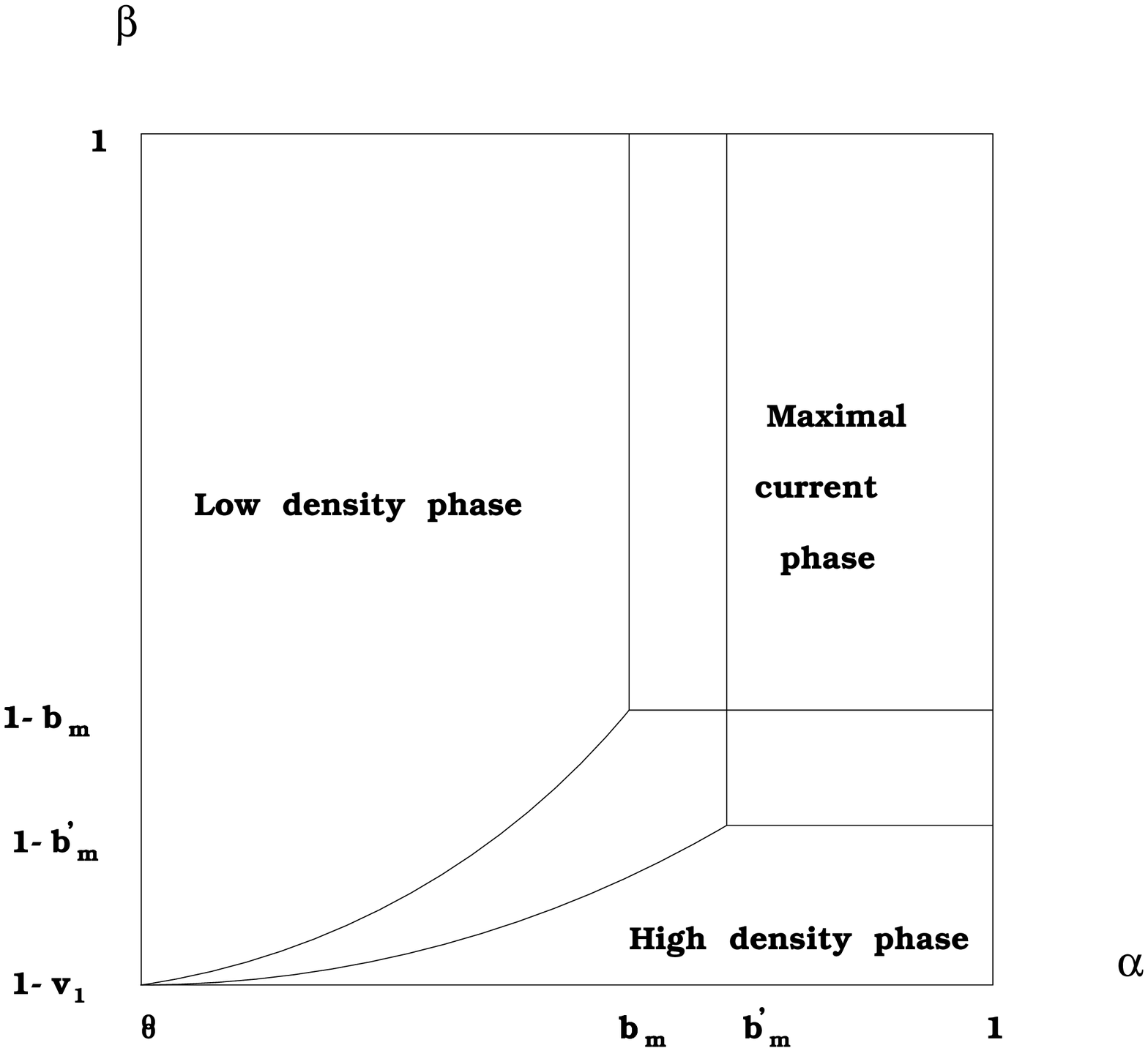}
\caption{Two phase diagram for the multi-species ASEP for different
distribution of hopping rates. In both cases $l[\sigma] < 0$ .} 
\end{figure}
${\bf \bullet}$ If on the other hand $l[\sigma]\geq 0 $, then the phase diagram consists of only two phases, namely the low
density and the maximum current phase. The extraction rate $\beta$ does not have any effect
on the system and only the injection rate $\a $ determines which
phase will develop in the system (fig.4). We also obtain the average
density of all types of particles in both phases.\\

\begin{figure}
\centering
\includegraphics[height=10cm,angle=00]{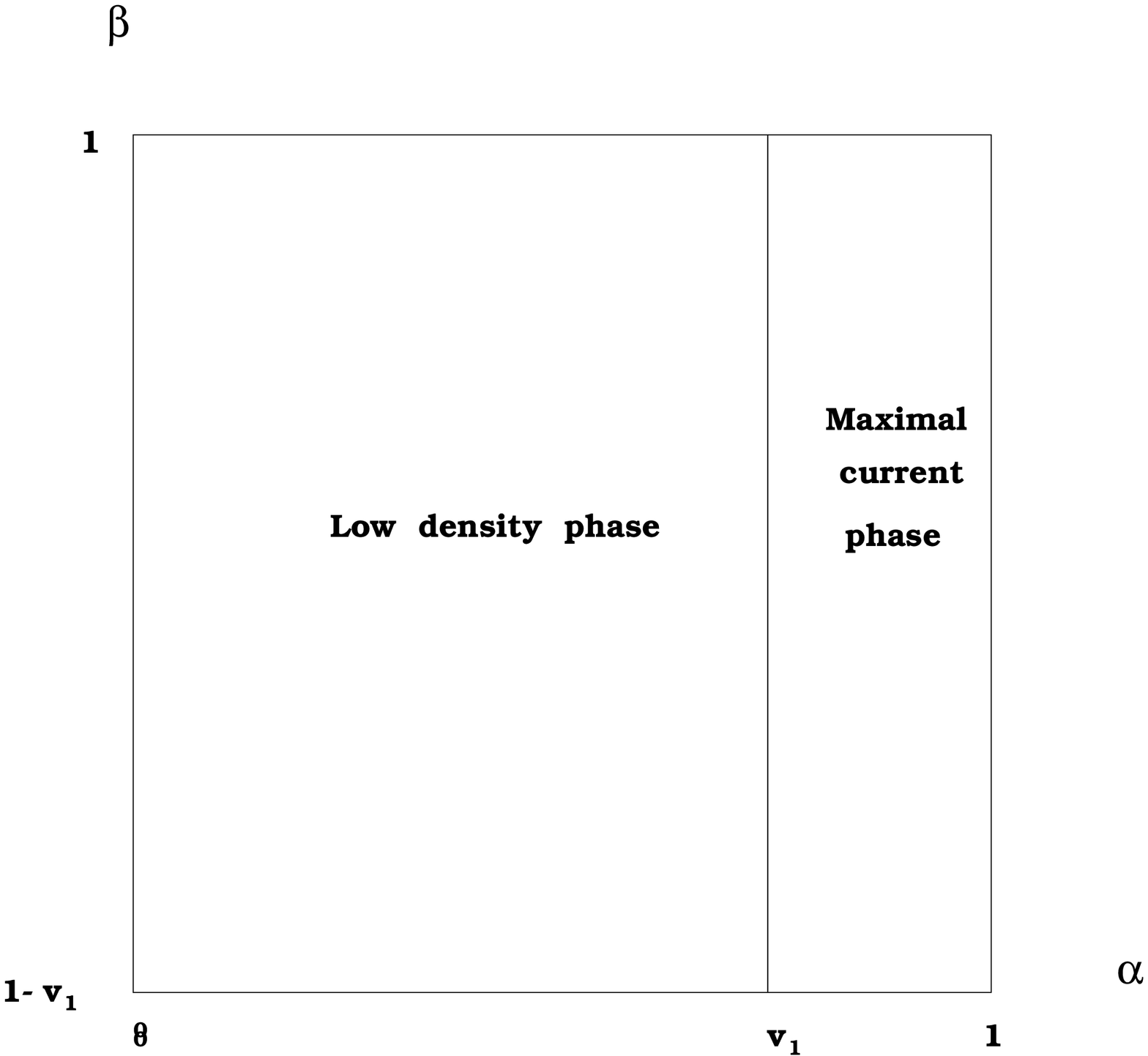}
\caption{Phase diagram of the multi-species ASEP when $l[\sigma]>0$}
\end{figure}

Thus the general shape of the phase diagram is controlled by 
three parameters. The two control parameters $\a$ and $\b$ represent the effect of
boundaries and the third parameter $l[\sigma]$, takes into account the distribution
function of hopping rates.\\
This paper is organized as follows: In section (2), we introduce the necessary
preliminary material from \cite{k1} and set our notations and conventions. In
section (3) the generating functions for the currents and average density
of particles of each species
are introduced and the former one is calculated exactly.
In section (4) the total
current is calculated in terms of which the different phases of the system
are determined. Section (5) is devoted to the calculation of the generating
function for the average densities and the calculation of
latter quantities.
In section (6) we discuss two special cases, namely the 1-ASEP where we
reproduce
the already known results, and the case when one of the hopping rates is
much smaller than the others. We conclude the paper with a discussion in
section (7) .
\section{Some algebraic preliminaries}
To make the present paper self-contained, we quote the basic
definitions and theorems from \cite {k1}, to which the reader can refer to,
for
further details and proofs.\\                      
The algebra (\rf {v4}-\rf{v4'''}) has only one or infinite dimensional representations.
We write the infinite dimensional representation in a convenient
basis consisting of vectors $ |0>, |1>, |2>, \cdots $, with the following actions:
\begin{eqnarray} E |n>&=& |n+1> \\
D(v)|n>&=& v^{-n}|0> + v^{-n+1}|1> + v^{-n+2}
|2> + ... v^{-1}|n-1> + |n> \end{eqnarray}
We also have:
\be |V> = \sum_{n=0}^{\infty} (1-\b)^n |n>\ee
\be <W| = \sum_{n=0}^{\infty} (\a)^{-n} <n|\ee
The basis$\{<n|\} $ is the dual of the basis $\{|n>\}$, i.e:
$<n|m>= \delta_{n,m}$.\\
The operator $ C $ is defined as
\be \la{C1} C:= E + \int \sigma(v) D(v) dv \ee
whose action on the basis vectors is calculated to be
\be \la{C2} C|n> = \sum_{k=0}^{n+1} <{1\over v^{n-k}}>|k>\ee
where
$<{1\over v^k}> $ is the average of the inverse $k-$th power of the hopping rates.
($<{1\over v^k}> = {1\over p}\sum_{j=1}^p {1\over (v_j)^k}$ for discrete distributions
and $<{1\over v^k}> = \int dv \sigma(v) {1\o v^k} $ for continuous distributions).
The explicit matrix form of $ C $ is :
\be \la{mat} C = \left( \begin{array}
{cccccccc}1&<{1\over v}>&<{1\over v^2}>&<{1\over v^3}>&<{1\over v^4}>&.&.&.\\1&1&<{1\over v}>&<{1\over v^2}>&<{1\over v^3}>&.&.&.\\
0&1&1&<{1\over v}>&<{1\over v^2}>&.&.&.\\0&0&1&1&<{1\over v}>&.&.&.\\0&0&0&1&1&.&.&.\\
.&.&.&.&.&.&.&.\\.&.&.&.&.&.&.&.\\.&.&.&.&.&.&.&. \end{array}\right). \ee
Clearly direct evaluation of the N-the power of this matrix is a formidable
task. In the one species case where $<{1\o {v^k}}> = 1,\ \ \forall k $, one can go to
a new basis $\{|n>':=(E-1)^n|0>\}$ in which the matrix $C$ becomes tri-diagonal
with the simple form $ C_{n,m} = 2\delta_{n,m}+\delta_{n,m+1}+\delta_{n,m-1}$.
This Hermitian matrix can then be either easily diagonalized \cite{dehp}
or else, yields simple recursion relations which can be solved by an analogy
with the master equation of a random walk
in the presence of an absorbing wall \cite{dehp}. 
Due to the complicated form of the matrix $C$, none of the above strategies
work in the present case.
The same is true with the method of repeated application of the algebraic
relations (\rf{v4}-\rf{v4'''})
and calculating directly the matrix element
$<W|C^N|V>$ \cite{dehp}.\\
There is however one basis in which a manageable
recursion relation can be found, namely the coherent basis defined as follows:
\be\la{ch1} \vert u > = \sum_{n=0}^{\infty} u^{n}\vert n > \ee
\be\la{ch2} < u \vert = \sum_{n=0}^{\infty} u^{n}<n\vert. \ee
Note a slight difference of our notation with that of \cite{k1} in the symbol
for $<u|$, where this state would have been denoted by $<u^{-1}|$.
These states have the following properties:
\be \la{w1}<u|E = u <u|\ee
\be \la{w2}D(v) \vert u > = { v\over { v - u}}\vert u > \ee
\be \la{w3}<\omega\vert u > = {1\over {1-{u\omega}}} \ \ \ {\rm  for } \
\ \ \ \vert u \omega \vert < 1 \ee
\be\la{w4}  \oint_{c} {d u \over {2 \pi i u} }\vert u><u^{-1}\vert = I \ee
where $ I$ is the identity operator and $c$ is any contour encircling the origin.\\
{\bf Remark}:  In calculation of matrix elements of operators between two states
$<a| $ and $|b>$, one can insert
any numbers of unit operators in the form of (\rf{w4})
with integration
variables
$ u_1, u_2, \cdots $ from left to right,
provided that $ |{1\o a}|> |u_1|> |u_2|>
\cdots |b| $. This is due to the restriction (\rf{w3}). The results of such calculations are then valid only for
$ |ab|<1$ and must be analytically  continued to larger domains.
From the definition of $C$ and (\rf{w1})-(\rf{w3}) one obtains
\be\la{N2} <u|C|w> = \Big( u+\int dv \sigma(v) {1\over {1-w/v}}
\Big) ({1\over {1-uw}})=: \Big(u+ g(w)\Big) ({1\over {1-uw}})\ee
where the second equality defines the function $ g(w)$, a shorthand and useful notation
for which is
\be \la{gb} g(w)= \left< {1\over {1-w/v}}\right>.\ee
Here the average is taken with respect to
the probability distribution of hopping rates, $\sigma(v)$.

\section{The generating functions for current and average densities}
The total current for a system consisting of $N$ sites has been found to
be \cite{k1}
\be\label{82}
J={{<a|C^{N-1}|b>}\over{<a|C^N|b>}},
\ee
where $<a| $ and $ |b>$ are coherent states and for convenience,
we have denoted $1-\b$ by $b$, and $\a^{-1}$ by $a$.
The current density of particles of velocity $ v $ is given by 
\be\la{82a} J(v) = v \sigma(v) J\ee
In the thermodynamic limit $N\to\infty$, there is a simple way to evaluate
(\rf{82}). Define a generating function
\be\label{81}
f(s;a,b):=\sum_{N=0}^\infty s^N<a|C^N|b>
\ee
The convergence radius of this formal series, $R$, is precisely what we
need. In fact
\be \la {rem}
R=\lim_{N\to\infty}{{<a|C^{N-1}|b>}\over{<a|C^N|b>}}.
\ee
{\bf Remark:} The function  $f(s;a,b) $
has a Taylor--series expansion in terms of (non-negative) powers of its three
arguments, which means that there is a region containing the origin of the
space $C^3$, where $f$ is analytic. This is easily seen by noting
that $ C^N$ can in principle be expanded in terms of integrals of monomials of
the form $ E^m D(v_1)D(v_2)...D(v_n) $ and using the relations
$ D(v)|b> = {v\o {v-b}}|b> $ and $ <a|E = a<a|$.\\

The radius of convergence is also the absolute value of the nearest
singularity of $f$ to the origin. We also know that all of the coefficients
of the Taylor expansion of $f$ in terms of $s$ are positive. This assures
that the nearest singularity of $f$ lies in fact on positive real half-line.
That is, the current at the thermodynamic limit is real and positive, as it
should be.
A similar method works for the average density of particles of each species as well. The global density of
particles of type $i$ is
\be
\rho(v)={\sigma(v) \over{N}}\sum_{k=1}^N{{<a|C^{k-1}{D(v)}C^{N-k}|b>}
            \over{<a|C^N|b>}}.
\ee
To evaluate this, we use fugacities $z_0 $ and $z(v)$ to define an operator $C[{\bf z}]$
as
\be
C[{\bf z}]:=z_0E+\int z(v)\sigma(v)D(v) dv .
\ee
Note that we have $C[{\bf 1}]=C$,
where by ${\bf 1}$ we mean the fugacities $ z_0 = z(v) = 1 $.
It is straightforward to see that
\be
\rho (v) = {z(v)\over N}{\delta\over{\delta z(v)}}
           \ln <a|C^N[{\bf z}]|b>\Big\vert_{{\bf z}={\bf 1}}.
\ee
Once again, the right--hand of this can be expressed is terms of the radius
of convergence $R({\bf z})$ of a formal series $ f({\bf z};s,a,b)$ defined as:
\be\label{85}
f({\bf z};s;a,b):=\sum_{N=0}^\infty s^N<a|C^N[{\bf z}]|b>.
\ee
Using an equivalent definition for the radius of convergence as
$R({\bf z}):=\lim_{N\longrightarrow \infty}\Big((<a|C^N[{\bf z}]|b>\Big)^
{-1\o N}$, we have:
\be\label{24}
\rho(v)=z(v){\delta\over{\delta z(v)}}\ln{1\over{R({\bf z})}}
           \Big\vert_{{\bf z}={\bf 1}}.
\ee
So the key step in obtaining the physical quantities is to calculate the
functions (\ref{81}) and (\ref{85}), which we call the generating functions
for currents and average densities respectively.

\sse {Exact calculation of the generating function $f(s;a,b)$}
First we use (\rf{N2}) to obtain a recursion relation for $<a|C^N|b>$:
\bea\label{51}
<a|C^{N+1}|b>&=&\oint{{\d u}\over{2\pi i u}}<a|C^N|u><{1\o u}|C|b>\cr
    &=&\oint{{\d u}\over{2\pi i u}}<a|C^N|u>\left[g(b)+{1\over u}\right]
       {1\over{1-b/u}},
\eea
where the function $ g$ has been defined in (\rf{gb}), 
i.e. $ g(b) = < {v\o {v-b}}>$.
Multiplying both sides of (\ref{51}) by $s^N$ and summing over $N$ from
zero to infinity, we arrive at
\be\label{55}
{1\over s}[f(s;a,b)-f(0;a,b)]=\oint{{\d u}\over{2\pi i u}}
      f(s;a,u)\left[g(b)+{1\over u}\right]{1\over{1-b/u}}.
\ee
The generating function, which we calculate in this way, will be restricted to
the domain $|ab|<1$. After calculating it for this region of
parameters, we will analytically continue it for other values of
parameters as well.   
Since $ |{1\o a}|>|u|>|b|$ the integrand
in the right--hand side of (\ref{55}) has just two poles inside the
integration contour; one at $u=0$ and the other at $u=b$. This is true
provided $f(s;a,u)$ itself is analytic for $u$ inside the integration
contour. However, we know that for small values of its arguments the function
$f(s;a,u) $ is analytic
(see the remark after eq. (\rf{rem})).
The result of this
calculation will be valid for small values of the arguments of the
generating function. One can then use analytic continuation to obtain
more general results. Knowing the non-analytic structure of the
integrand, one can use Cauchy's theorem to evaluate the right--hand side of
(\ref{55}):
\be\label{56}
{1\over s}[f(s;a,b)-f(0;a,b)]=-{{f(s;a,0)}\over b}+
\left[{1\over b}+g(b)\right]f(s;a,b).
\ee
Solution of this equation for $f(s;a,b)$ yields
\be\label{57}
f(s;a,b)={{sf(s;a,0)-b\; f(0;a,b)}\over{b\{s[g(b)+1/b]-1\}}}
\ee
Note that from (\rf{81})
\be\label{512}
f(0;a,b)=<a|b>={1\over{1-ab}},
\ee
and $g(b)\equiv \Big<{1\o {1-{b\o v}}}\Big> $ is a function
which can be determined once the data of the problem
(i.e: the distribution function $P(v)$ ) are given. Equation (\rf{57}) then suggests
that $f(s;a,b)$ is known, provided a two--variable restriction of
it, namely $f(s;a,0)$ is known. Eq.(\ref{57}) contains even more information. To
see this, notice that from (\ref{57}) it seems that there is a pole for $s$
at 
\be\la{s0}
s_0=S(b):={1\over{g(b)+1/b}}
\ee
From the definition of $g(b)$, it is seen that, as $b$ tends to zero, $g(b)$
tends to unity.
So, for small values of
$b$, $S(b)$ behaves like $b$:
\be
S(b)\sim b+O(b^2),\qquad {\hbox{\rm as}}\; b\to 0.
\ee
But this means that as $b$ tends zero, the radius of convergence for the
variable $s$ tends to zero, and this can not be the case, due to the remark after
(\rf{81}).
To avoid this apparent
paradox, it must be true that $s=S(b)$ must not really be a pole, at least
for small values of $b$. This means that the numerator in (\ref{57}) must
also vanish for $s=S(b)$. For this to be the case, we should have
\be\label{58}
S(b)\; f[S(b);a,0]=b\; f(0;a,b)= {b\over 1-ab}.
\ee
This equation allows us to determine the function $f(s;a,0)$ and hence via
(\rf {57}), the complete generating function. Denoting the
inverse function of $S$
by $B$:
\be\label{59}
S[B(s)]=s,
\ee
we find
\be\label{510}
f(s;a,0)={1\o s}{B(s)\o 1-aB(s)}.
\ee
Note, however, that $S$ is not in general one to one
and in different domains of its arguments it has different inverses.
In fact we will show later that $ S $ is at most a two-to-one function
with the property $ S(0)= 0 $. By the inverse $ B $ we mean the one
that tends to zero as its argument tends
to zero:
\be
\lim_{s\to 0}B(s)=0.
\ee
Inserting (\ref{510}) in (\ref{57}), and using (\rf{58}), we find
\be\label{513}
f(s;a,b)={{\displaystyle{B(s)\over{1-a\; B(s)}}-{b\over{1-a\; b}}}
\over{\displaystyle b\left[{s\over{S(b)}}-1\right]}}.
\ee
This is the final form of the generating function. For any given probability
distribution of hopping rates, one can obtain $ S(b)$ and hence $B(s)$ from (\rf{gb}), 
and (\rf{s0}), which after insertion into (\rf{513}) gives the complete generating function.
What we will do in the next sections is to carry out an analysis of the
singularity structure of this function and determine the currents and hence
the different phases of our multi-species stochastic process. Our results
and analysis depend on the general behavior of the functions $ S(b)$
and $ B(s)$ which in turn depend on the distribution of hopping rates.

\section{The total current and the phase structure of the system}
\sse{Properties of the function S}
As it was seen in the previous section, to investigate the properties of the
system, one must know the behavior of the function $S$. We have
\be\label{80}
{1\over{S(b)}}={1\over b}+g(b),
\ee
from which we find
$${d^2\o db^2}S^{-1}(b)= {2\o b^3} + \Big<{ 2
v\over (v-b)^3}\Big> > 0. $$\\
Thus $ S^{-1}(b) $ is a concave function. Combination of this with the fact that
this function is positive for $ b\in (0, v_1)$ and tends to infinity as
$ b\longrightarrow 0 $, implies that $ S^{-1}$ has at most one minimum in $(0,v_1)$.
Hence $ S $ is a positive function in $(0,v_1) $ and has at most one maximum
in this domain. Note also that $ S(0) = 0 $. \\
The phase structure depends crucially on whether $ S $ attains a local maximum
in this domain (i.e.$(0, v_1)$) or not. This is easily checked from the
sign of $ S'(v_1^-) = 
 {1\o {v_1^2}} - \Big<{v\o{(v-v_1)^2}}\Big> $.
This quantity is determined {\it only by the probability distribution of hopping
rates} and this is where this function plays its essential role. 
To emphasize the dependence on
the distribution function we denote this quantity by $l[\sigma]$\\
\be \la{dis} l[\sigma]:= {1\o {v_1^2}} - \Big<{v\o{(v-v_1)^2}}\Big> .\ee
As we will see if $l[\sigma]<0$ there are three regions in the phase diagram
namely the high-density, the low-density and the maximum current phases.
On the other hand if
$l[\sigma]\geq 0$, the high density phase disappears and only the low-density and the
maximum current phases remain. We call these two regimes, the 3-phase
and 2-phase regimes respectively.
Qualitatively the transition from the 3-phase
to the 2-phase regime is accomplished by shifting the distribution function
from low speeds to higher speeds. As an example if $\sigma(v_1)\ne0$( i.e:
if there is a significant relative probability of injecting slow particles 
to the system), then it is clear that $l[\sigma] = -\infty$, which means that we are in the 3-phase regime.
The case
of discrete values of particle velocities is a special case in this
category. However if $\sigma(v)$ approaches zero slowly enough as $ v $ approaches
$v_1$, (i.e: if the chance of entrance of slow cars is small), then we will
be in the two-phase regime. The exact criterion is given by the parameter
$l[\sigma_c]=0$.\\
In the sequel we will need one further property of $ S $. From its definition
it is seen that $ S $ as a function of the complex variable $b$
has a singularity at $ b= v_1 $. If the distribution is discrete, this singularity is
a simple pole. If the distribution is continuous, $ S $ has a branch cut
on a
segment of the real line beginning from $b=v_1$ to $ +\infty$. To see this we use
 $ {1\o {x\pm i\epsilon}}= pf({1\o x})\mp i\pi\delta(x)$
to obtain
$$ {1\o S(b+i\epsilon)}-{1\o S(b-i\epsilon)}= 2\pi i b \sigma(b).$$

\sse{The singularities of the generating function $f(s;a,b)$}
In order to determine the phases we have to determine the singularities of the
generating function.
For the
case $l[\sigma]<0$ where $ S $ has a maximum $ s_m $ at $ b_m \in (0,v_1) $ (Fig. 5),
we define two right inverses for $S$.
\begin{figure}
\centering
\includegraphics[height=10cm,angle=00]{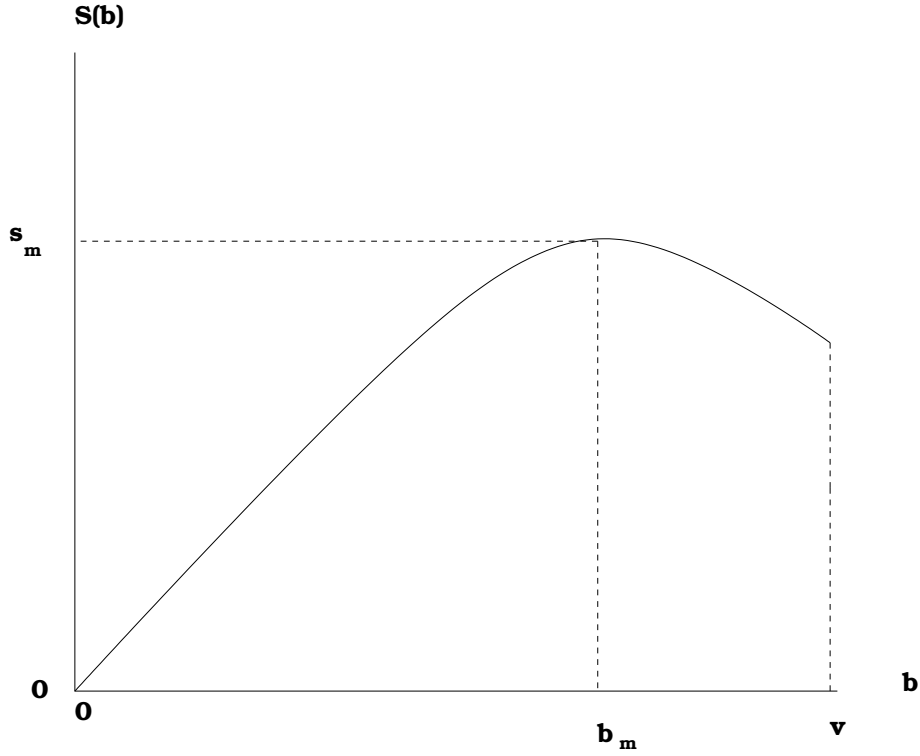}
\caption{The generic form of the function $S(b)$ that produces the three phase regime}
\end{figure}

One is
the function $B$ defined in the previous section. It is defined in the
interval $[0,s_m]$, and has the following properties.
\be\label{86}
\cases{S[B(s)]=s,&$0\leq s\leq s_m$\cr
       B[S(b)]=b,&$0\leq b\leq b_m$\cr}.
\ee
The other function is $\tilde B$, defined in $[S(v_1),s_m]$, with the
following properties
\be\label{87}
\cases{S[\tilde B(s)]=s,&$S(v_1)\leq s\leq s_m$\cr
       \tilde B[S(b)]=b,&$b_m\leq b\leq v_1$\cr}.
\ee
The graphs of these functions are shown in figure 6.
\begin{figure}
\centering
\includegraphics[height=10cm,angle=00]{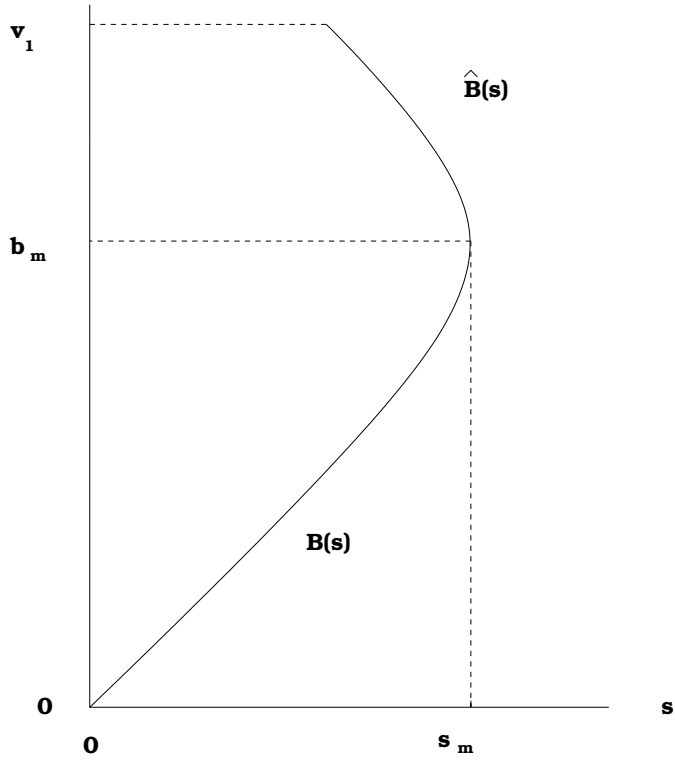}
\caption{The functions $B(s)$ and $ {\tilde B}(s)$}
\end{figure}

%\begin{picture}(0,0)(7,7)
%\put(0,0){\special{em:g5.bmp}
%\end{picture}
We now consider the
generating function as given in 
(\ref{513}). 
The singularities of $f$ as a function of $s$ may be of one of following
types.
The first singularity denoted be $ s_{\a}$ may arise from vanishing of
$(1-aB(s))$ in the numerator. That is
\be\label{89}
B(s_\alpha)={1\over a}=\alpha,
\ee
or
\be\label{810}
s_\alpha =S(\alpha).
\ee
However (\ref{89}) has a real positive solution for $s$ if and only if
$\alpha=(1/a)<b_m$ (see fig.(6)).
Thus $s_\alpha$ is a singularity provided that $\alpha<b_m$.\\
The second singularity which we denote by $ s_{\b}$ may arise from
vanishing of the denominator, i.e:
\be\label{88}
s_\beta =S(b)=S(1-\beta ).
\ee
However for $b< b_m$, according to (\ref{86}) the numerator vanishes
as well, and $s_{\b}$ is no longer a singularity, which means that
$s_{\b}$ is a singularity only when $b>b_m$ (or $\beta <1-b_m$).\\
Finally, $B$ itself becomes singular at $s=s_m$. Notice that $s_m$ is
greater than $s_\alpha$ and $s_\beta$ both, provided these latter two exist
(see fig.(5)).\\
In the next subsection we use these information to determine the phases.
\sse{The phase structure for $l<0$}
In this case, the generating function has three singular points, namely
$s_{\a},s_{\b}$ and $s_m$,
each phase (the analytic expression of the current) is determined according to
which of these singular points are the smallest.
\begin{itemize}
\item{{\bf The low-density phase ($s_{\a} < s_{\b} , s_m $):}}\\
In this phase which develops when $ \a < b_m $ and $ S(\a) < S(1-\b),$
we have:
\be\label{j53}
J=S(\alpha),
\ee
\item {{\bf The high-density phase ($s_{\b} < s_{\a} , s_m $):}}\\
In this phase which develops when $\b < 1-b_m$ and $ S(1-\b)> S(\a),$
the total current is given by 
\be
J=S(1-\beta),
\ee
and finally
\item{{\bf The maximum current phase ($s_m$ is the only singularity):}}
This phase exists for the  rectangular area $ \a > b_m $ and $ \b > 1-b_m $.
The current is given by
\be
J=s_m.
\ee
\end{itemize}
Summarizing we have:
\be \la{sum} J = \cases
{ S(\a),& $\ \ \ \a<b_m,\ \ \ {\rm and}\ \ \  S(\a)<S(1-\b)$\cr
       S(\b), &$\ \ \ \b< 1-b_m,\ \ \ {\rm and}\ \ \ S(1-\b)<S(\a)$\cr
       s_m = S(b_m), &$\ \ \ \ \a> b_m,\ \ \ {\rm and}\ \ \ \b> 1-b_m $ \cr}.\ee
The phase diagram is shown in fig.(2).
The coexistence curve between the low and high-density phases is
obtained by the nontrivial solution of the equation $S(\a) = S(1-\b)$.
A parametric representation of this curve is:
\be 
\cases
{ \a = B(s)\ \ \ \ \cr \ \ \ \ \ \ & $ S(v_1)< s < s_m $
 \cr \b = 1-{\tilde B}(s)
\cr}\ee
We see that the main features of the
1-ASEP diagram is present. In this regime the multi-species nature
of the process has only a minor effect. We will see that this exact result is also
substantiated by a domain wall analysis in accordance with the analysis of \cite
{schal}. 
\sse{The phase structure for $l\geq 0$}
In this case the generating function has only two singular points, namely $ s_{\a}$
and $s_m = S(v_1)$. Consequently we have only two phases (Fig. (4)).\\
$ {{\bf \bullet}}$ The low density
phase in which
\be
J=S(\alpha),
\ee
and\\
$ {\bf \bullet}$ The maximum current phase in which
\be
J=S(v_1).
\ee
In summary
\be \la{summ} J = \cases
{ S(\a),&$\ \ \ \ \a\leq v_1 $\cr
       S(v_1), &$\ \ \ \ \a> v_1$\cr}\ee
The high--density phase has been shrunk and lost, and the injection parameter
determines which phase we have.
This is due to the fact that in this regime the current density diagram
is monotonically increasing 
(see section 7) and hence according to the domain wall analysis
, only the low density and the maximum current phases
are expected to exist. Moreover in the maximum current phase everything is controlled
by the lowest-speed particles.\\
The disappearance of the maximum density
phase has an interesting implication, for the occurrence of traffic jams and
its dependence on the bulk parameters beside the boundary ones.
From the above analysis one can conclude that the maximum density
or traffic jam occurs
only when there is a critical probability of having particles or cars
of slow velocities, the exact criteria is given by the parameter $l$ defined
above.
\section{Exact calculation of the generating function $ f({\bf z};s;a,b)$
and the average densities of each species}

Using the same calculation, which led to (\ref{55}), we obtain
\be\label{20}
{1\over s}[f({\bf z};s;a,b)-f(0;a,b)]=\oint{{\d u}\over{2\pi i u}}
      f({\bf z};s;a,u)\left[g({\bf z}; b)+{z_0\over u}\right]{1\over{1-b/u}},
\ee
and from that we arrive at an expression analogous to (\ref{57})
\be\label{21}
f({\bf z}; s;a,b)={{z_0\; s\; f({\bf z}; s;a,0)-b\; f({\bf z};0;a,b)}\over
{b\{s[g({\bf z}; b)+z_0/b]-1\}}}
\ee
where
\be\label{g_1}
g({\bf z};b):=\left<{z\over{1-b/v}}\right> =
\int\d v\;\sigma(v){z(v)\over{1-b/v}}.
\ee
From (\ref{21}), by a reasoning exactly the same as that of section 2, we
arrive at
\be\label{22}
f({\bf z};s;a,b)={{\displaystyle{B({\bf z};s)\over{1-a\; B({\bf z};s)}}-
{b\over{1-a\; b}}}
\over{\displaystyle b\left[{s\over{S({\bf z};b)}}-1\right]}},
\ee
where
\be\la{g_2}
{1\over{S({\bf z};b)}}:={z_0\over b}+g({\bf z};b),
\ee
and $B({\bf z};s)$ is that right--inverse of $S({\bf z};b)$ which tends to
zero as $s\to 0$.

According to (\ref{24}), or its analogue for the case of continuous
distribution, the average density of particles of speeds between
$v$ and $v+dv$.
\be\la{g_3}
\rho (v)=z(v){\delta\over{\delta z(v)}}\ln{1\over{R({\bf z})}}
           \Big\vert_{{\bf z}={\bf 1}}.
\ee
Knowing the smallest singularity of
$f({\bf z}; s; a,b)$ is sufficient to obtain the average densities
$\rho (v)$. Once again, we can distinguish three phases: the low--density
phase, the high--density phase, and the maximum--current phase.
In the low-density phase $ R({\bf z}) = S({\bf z}, \a)$. Thus $ \rho(v)$
is obtained from (\ref{g_1},\ref{g_2}) and (\ref{g_3}) as follows:
\be \rho(v) = z(v)
{\delta\over{\delta z(v)}}
ln \Big [{z_0\over{\a}}+ \int\d v\;\sigma(v){z(v)\over{1-\a/v}}\Big]\Big\vert
_{{\bf z}=1} = {\sigma(v)S(\a)\over {1-{\a\over v}}}.\ee
The expression for the high--density phase is similar and reads
\be
\rho (v)={{\sigma(v)S(1-\beta )}\over{1-(1-\beta)/v}}.
\ee
The treatment of the maximum current phase however requires more care.
In the two-phase regime (i.e: $ l\geq 0$), 
where the function
$S$ does not attain any maximum in $[0,v_1]$, the maximum current is
$ S(v_1)$, the situation is the same as
above, and we have
\be
\rho (v)={\sigma(v)S(v_1)\over{1-v_1/v}}.
\ee
In the 3-phase regime (i.e: $l<0$) however, the maximum current is $S(b_m)$,
where $b_m$ itself depends on ${\bf z}$. Therefore,
\be
{\delta\over{\delta z(v)}}{1\over{S[{\bf z},b_m({\bf z})]}}
\Big\vert_{{\bf z}={\bf 1}}=
{\delta\over{\delta z(v)}}{1\over{S({\bf z},b_m)}}
\Big\vert_{{\bf z}={\bf 1}}+
{{\delta b_m({\bf z})}\over{\delta z(v)}}\Big\vert_{{\bf z}={\bf 1}}
{\partial\over{\partial b}}{1\over{S(b)}}\Big\vert_{b=b_m}.
\ee
The second term in the right--hand side is, however, zero, since ${1\over S}$ is
minimum at $b_m$. So we arrive at the expression
\be
\rho (v)={{\sigma(v)S(b_m)}\over{1-b_m/v}}.
\ee
To summarize, we have
\be\label{j51}
\rho (v)={{\sigma(v)S(x)}\over{1-x/v}},
\ee
where
\be\label{j50}
x=\cases{\alpha ,& low--density phase\cr
         1-\beta ,& high--density phase\cr
         b_m,& maximum current phase in the three--phase system\cr
         v_1,& maximum current phase in the two--phase system\cr}.
\ee
The average density of vacant sites $\rho_0$ can also be obtained either
by using the formula $ 
\rho_0 = z_0{\partial \over{\partial z_i}}\ln{1\over{R({\bf z})}}
           \Big\vert_{{\bf z}={\bf 1}} $ or by using the sum rule
\be\label{j52}
\rho_0+\int\d v\;\rho (v)=1.
\ee
From (\ref{j52}), one obtains for each phase determined by the parameter $ x$
defined in (\ref{j50})
\be \label{vac}
\rho_0 = 1-\int\d v{{\sigma(v)S(x)}\over{1-x/v}}.
= 1 - S(x)g(x)
\ee
where $S(x)=S({\bf 1},x) $ and $ g(x)=g({\bf 1},x)$. After using
(\ref{g_2}), this gives
\be\la{hole} \rho_0 = 1 - S(x)\big({1\over S(x)}-{1\over x}\big) = {S(x)\over x}.\ee

\section{Examples}
\sse{The single species ASEP}
In this case $ \sigma(v) = \delta (v-1)$. All we need to know to treat this special case is the function $S(b)$ as given
in eq.(\rf {80}). 
From (\rf{s0}) we find
\be S(b) = {1\o {b(1-b)}}\ee
with $ l=0, b_m = {1\o2}$, and $s_m= {1\o 4}$. Thus according to (\rf{sum}) we have
the following phases in accordance with previous results :
\be J = \cases{\a(1-\a) & $\ \ \ \a<\b,\ \ \ \ {\rm and}\ \ \ \ \a< {1\o2} $\cr
       \b(1-\b)& $\ \ \ \a>{1\o 2},\ \ \ \ {\rm and}\ \ \ \ \b< {1\o 2} $\cr {1\o 4}
       & $\ \ \ \a\geq {1\o 2},\ \ \ \ {\rm and}\ \ \ \b\geq {1\o 2} $\cr}.\ee\\
\sse{Continuous distributions; concrete examples of the disappearance of the maximum density phase}
In this section we consider two classes of distribution 
functions to see concretely the transition between the two and three phase 
regimes. Both of the distributions must be such that they vanish at $v_1$, 
otherwise as we have already remarked $ l=-\infty $ and we have three phases. .
For convenience we also rescale the time so that the average velocity is
no longer equal to unity.
Correspondingly the expression (\rf{dis}) for $l[\sigma]$ is replaced
by
\be
l[\sigma] = ({{\bar v}\over v_1})^2 - 
\Big<{v {\bar v}\o {v-v_1}^2}\Big>,
\ee
where ${\bar v}$ is the average hopping rate. \\
The first distribution that we study has a finite support.
$$\sigma(v) = A_m (v-v_1)^m ,\ \ \  m>0 ,\ \ \  v_1\leq v\leq v_2.$$
This kind of distribution has also been considered in a related context by Evans \cite{ev}.
Note that $m$ need not be an integer, and $A_m$ is a normalization constant to
be determined shortly. It is convenient to evalute the following integrals:
\be \la{int1}
I_k := \int_{v_1}^{v_2} (v-v_1)^k dv =
\cases{+\infty &$\ \ \ k\leq-1$\cr
       {\tilde I}_k := {(v_2 - v_1)^{k+1}\over {k+1}} &$\ \ \ -1< k $\cr}.
\ee
Obviously $ A_m = {1\o {\tilde I}_m}$. Simple calculations also yeild:
\be\la {vbar}
{\bar v} = v_1 + {m+1\o {m+2}}(v_2 - v_1) 
\ee
and 
\be\la{l1}
l[\sigma]= ({{\bar v}\o v_1})^2 - {{\bar v}\o {\tilde I}_m}(I_{m-1} + v_1 I_{m-2}),
\ee
from which we find
\be\la {ll1}
l[\sigma] = \cases{ -\infty &$ 0<m\leq 1$\cr
l({v_2\o v_1}, m) = ({{\bar v}\o v_1})^2 - (m+1){\bar v}\Big({1\o {m(v_2 -v_1)}} +  
{v_1\o {(m-1)(v_2 -v_1)^2}}\Big)&$ 1<m$\cr}.
\ee
The sign of the parameter $l(v_1, v_2, m)$ determines if the maximum density 
phase exists or not. In an extreme case the analysis of this quantity is
quite simple. For very large $m$, we find after inserting (\rf{vbar}) in (\rf{ll1})

\be\la {lll1}
l(x:={v_2\o v_1}, m)=  x^2 (1- {1\o {x^2-1}}) - {2\o m}x(x-1) + O(m^{-2}).
\ee
which implies that to zeroth order, if $x>2$ (i.e. $ v_2> 2 v_1 $), there is no maximum-density-phase. 
To first order the above condition is modified to $ v_2 > 2 v_1 (1+{1\o m}) $.\\
We now consider another distribution with infinite support.\\
$$\sigma(v) = A_m (v-v_1)^m 
e^{{-(v-v_1)\over \l}}, \hskip 1cm  m>0,\hskip 1cm  v_1\leq v $$
The analysis is similar to the previous case. We have
\be\la{l2} 
J_k := \int_{v_1}^{\infty} (v-v_1)^k e^{-{v-v_1\o {\l}}}dv = 
\cases{ +\infty &$ k\leq -1$\cr {\tilde J}_k := \l^{k+1}\Gamma(k+1) &$
-1<k$ \cr}
\ee
from which we find
\be\la{ll2}
{\bar v} = v_1 + m\l  
\ee
\be \la {lll2}
l[\sigma] = ({{\bar v}\o v_1})^2 - {{\bar v}\o {\tilde J}_m}(J_{m-1} + v_1 J_{m-2}).               
\ee
or 
\be\la{llll2}
l[\sigma] = \cases{ -\infty &$ 0<m\leq 1$\cr
l(x:={m\l \o v_1}, m) &$ 1<m$\cr}.
\ee
where 
\be
l(x, m) = x(2+x)-{1\o x^2}\Bigg({m+(2m-1)x\o {m-1}}\Bigg).
\ee.
Again to zeroth order of ${1\o m}$, we find
\be 
l = (x-{1\o x})(x+{1\o x}+2)
\ee
which implies that
when $ {\l m\o v_1} > 1 $, the high-density-phase disappears.

\sse {The p-species ASEP, with a hopping rate much lower than the others:}
The case of a fixed or moving impurity has been studied in many previous works
as for example in \cite{jl,djls,m}. In the present framework we can
consider a new case  where  the number of impurities is not one or even
fixed. That is we allow very slow particles to have a  chance of entering
into and leaving the system. That is we take $$ \sigma(v) = {1\o p} \sum_{i=1}^p
\delta(v-v_i)$$ 
and let one of the particles has a speed much lower than the rest: that is :
$ b< v_1<<< v_2<v_3<\cdots <v_p$. Then with the approximation $ 1-{b
\o v_i} \approx 1,\ \ {\rm for}\ \ i=2,3,\cdots p $, 
one can
write:
\be g(b)\equiv {1\o p}\sum_{i=1}^{p} {1\o{1-{b\o v_i}}} \approx
{1\o p}\Big( {1\o{1-{b\o v_1}}}+p-1\Big) \ee
from which one obtains:
\be\la{ce}  S^{-1}(b) \equiv {1\o b} + g(b) = {b+1\o b}+{b\o {p(v_1-b)}}\ee
For this function we have:
\be b_m = {v_1\o{1+\sqrt{{v_1\o p}}}} \ \ \ \ {\rm and} \ \ \ \  
 s_m = {v_1\o{1+ v_1 + 2\sqrt{{v_1\o p}}}} \ee
Besides $v_1, \a $ and $ \b$,
only the number of species plays a role here. 

The phase diagram is shown in Fig.(7).
\begin{figure}
\centering
\includegraphics[height=5.5cm,angle=00]{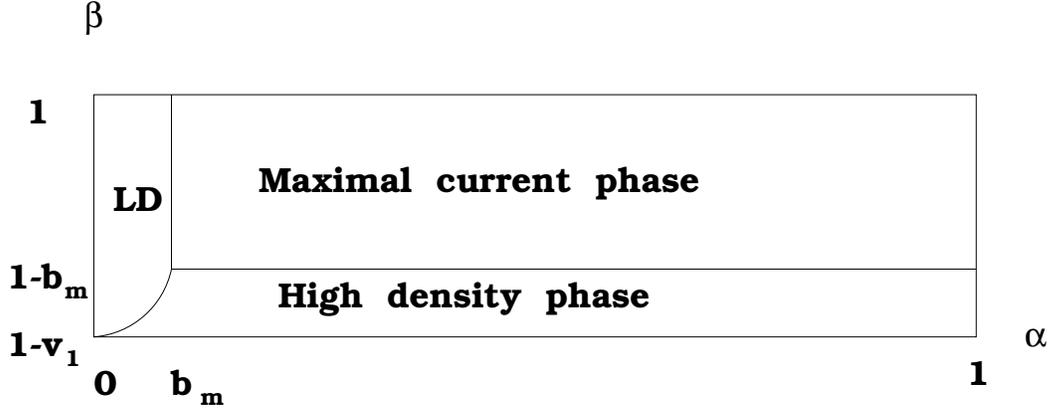}
\caption{Phase diagram of the p-species ASEP
when one of the
hopping rates is
much smaller than the others.}
\end{figure}

The following features are readily
observed. Compared to the other two phases, the size of the low density
region is very small, as we expect on physical ground. Only for very small
injection rate $\a $ and for very large extraction rate $\b$ can this phase
exist in the system. Even in the maximal current phase the current which is
given by $ s_m $ is seen to be small and limited by the speed of the lowest
particles. For fixed \ \ $\a,\ \ \ \b,\ \ \ {\rm and }\ \ \  v_1 $,\ \ \ as
the number of species $p$
increases, the value of
$b_m $ approaches $v_1$ and hence the high density phase begins to shrink,
leaving only two phases in the system.\\
It is instructive to calculate the relative numbers of particles of different
types including holes in each of the above phases.
From (\rf{j51}) we find
\be  {\rho(v_ i)\o \rho(v_1)} = {1-{x\o v_1}\o {1-{x\o v_i}}}\ee
where in each phase $x$ is given as in (\rf{j50}).
Inserting the relevant values of $x$ and in the approximation ${b\o v_i}
<<1\ \ \ \forall i\geq 2 $,  we find:
\be {\rho(v_i)\o \rho(v_1)}= \cases{ {1-{\a\o v_1}\o {1-{\a\o v_i}}},
& Low--density phase\cr 1-{b\o v_1}, & High--density phase
\cr 1-{b_m\o v_1}, & maximal current phase\cr}\ee
In the low density region where 
$\a$ takes values from $0$ to $ b_m $, this ratio takes values from
$1$ to $1-{b_m\o v_1}= 1-{1\o {1+\sqrt{{v_1\o p}}}}$. Thus in this region
almost all types of particles are present in the system. In the high
density region this ratio is  at most equal to $1-{1\o{1+\sqrt{{v_1\o p}}}}$
which is determined by the interplay of the lowest speed and the
number of species. For large $p$, it is indeed a very small value ,
indicating that the system is
almost filled by the lowest particles

This ratio takes its maximum value in the entire maximal current phase.
One can also obtain the ratio of the density of lowest speed particles
to that of the holes. From (\rf{j51}) and (\rf{hole}) one obtains 
\be {\rho(v_1)\o \rho_0}=
\cases{{1\o p}{\a\o {1-{\a\o v_1}}},\cr {1\o p}{b\o {1-{b\o v_1}}}\cr 
{1\o p}{b_m\o {1-{b_m\o v_1}}}\cr}\ee
where from top to bottom we have listed the low density, the
high density and the maximal current phases. Again we can find the
limiting values of this ratio, to see how crowded the system is, in each
phase. It is simple to see that the ratio ranges from $0$ to
$\sqrt{{v_1\o p}}$ in the low density, and from
$\sqrt{{v_1\o p}} $ to infinity, in the high density and is fixed at
$\sqrt{{v_1\o p}}$ in the maximal current phase.\\
With a little more effort, starting from (\rf{ce}), the coexistence line
between
the low-density and the high-density phases is found to be given by:
\be \b = \a + 1 - {pv_1^2 + {\a}^2(v_1-p)\o (\a+p)v_1 - \a p }\ee

\se{ Discussion}
The results that we have obtained on the phases and currents are exact.
We can get a feeling for these results, based on the intuitive arguments of domain wall dynamics \cite{sph,abl,jl, schd,schal}.
What we will do in this section is to formally adopt the analysis of  \cite{schal} and
redrive our exact results.
The essential result of \cite{schal} is that for all single species 
processes
which have single peak current density relation, the phase diagram of the
ASEP is generic, that is the possible phases are the low density,
the high density
and the maximal current phases. Roughly speaking one expects that for
$ \a$ small and $\b$ large, the low density phase denoted schematically by
(000000000), prevails in the system, and
for $\a$ large and $\b$ small, the high density phase denoted by (11111111)
prevails. However when there
is no restriction on the injection and extraction rates of the particles,
that is for $\a\ \ \  {\rm  and }\ \ \  \b$ large, the current reaches
its maximum value allowed in the current density diagram, this new phase
being called the maximal current phases and  denoted by
(mmmmmmmm). The exact shape of the phase diagram and the coexistence
lines
are obtained by studying the dynamics of a supposedly formed domain wall at
sufficiently late times
between any pair of these phases under appropriate conditions. For example
when $\a$ is small and $\b$ is large, the late time configuration is supposed
to be (000000111111). It is also assumed that deep into each of the two
segments
we have a product measure with constant density. This
assumption is well-founded \cite{sph,abl,jl,schd}
by numerical, mean
field and exact solutions .
The velocity of such a domain wall is then given by the formula 
\be\label{111} V = {J_o - J_1\o {\rho_0 - \rho_1}},\ee
where $J_0$ and $ \rho_0$ ( resp. $J_1$ and $ \rho_1$ ) are the current and
density to the far left ( resp. right ) of the domain wall.
The sign of this velocity determines the prevailing phase and setting this
velocity
equal to zero determines the coexistence line. In the latter case the two
phases coexist due to dominance of fluctuations in the rms position of the 
domain wall. For the currents and the densities in (\rf{111}) one uses
the mean field values. For the maximal phase, one uses the density which
maximizes the current in the current density diagram.\\
The above analysis can be readily applied to the multi-species case.
On the
assumption that the coarse grained bulk current is given by the uncorrelated
Bernoulli measure \cite{lps}, we can use the one-parameter family of
one dimensional representations for
the bulk relations in (\rf{v4}-\rf{v4'}) to obtain \cite{k1}
\be E= {1\o b}\hskip 1cm D(v) = {v \o{v - b}} \ee
and consequently the following forms for current and total density: 
\bea \la{1d}J(b) &=& C^{-1} =  
\Big ( {1\o b} + \Big< {v\o {v-b}}\Big>\Big)^{-1}\\ \la{1d'}
\rho(b) &=& \Big<{v\o{v-b}}\Big>
\Big ( {1\o b} + \Big< {v\o {v-b}}\Big>\Big)^{-1}\eea
Note that the right hand side of (\rf{1d}) is exactly the function $ S(b)$
defined in equation (\rf{s0}). However, before using equation (\rf{111}), we need 
to determine the free parameter $b$ and its range, in the Bernoulli measure
for each phase.
For the maximum current phase the parameter is obviously $ b_m$ which
maximizes
$ J(b)$. This is exactly the parameter, which has been defined in section
(4.1).
For the other two boundary-controlled phases, the parameter $b$ should be
fixed by compatibility with the boundary conditions of (\rf{v4''}-\rf{v4'''}),
according to which a product measure coupled to a left reservoir injecting
particles at rate $\a$
should have $ b={\a} $ and a product measure coupled to a right reservoir
extracting particles at rate $\b$ should have $ b={1-\b}$.
Instead of using this type of argument which is based on MPA relations
one can follow the more general argument suggested in \cite{schd},
to match the boundary rates with the bulk densities.\\
Note also that for the low density phase $b<b_m$ and for the high density
phase $b>b_m$.
Thus we have
$ J_0 = J(b=\a), \ \ \  J_1 = J(b=1-\b)\ \ \  {\rm and }\ \ \ J_m = J(b_m)$.\  
Equating the currents we obtain exactly the phase structure previously obtained
by exact solution. Moreover the size of the maximum density region in the
phase diagram depends on the value $v_1-b_m$ (see Fig. (5)).
When $v_1 - b_m $ approaches zero, the size of this region shrinks
and we remain only with two phases. This is again in accord with our exact
solution. \\
To conform completely to the picture advocated in \cite{schal} we should have
discussed various phases according to the behavior of the function
$J(\rho)$ and not the function $J(b)$. However in our case the qualitative
behavior of these two functions resembles each other. In fact it is seen from
(\rf{1d}-\rf{1d'}) that
$\rho$ is a monotonically increasing function of $b$,
which attains its maximum $\rho_1$ at $b=v_1$.
Moreover
\ $ J(\rho=0)=0,\ \  J'(\rho=0)=1$,\ 
and finally \cite{k2} $ J $ is a convex function of $\rho$.
To see if $J(\rho) $ attains a local maximum in its domain of definition
$[0,\rho_1]$ or
not we evaluate ${dJ\o {d\rho}}$ at $\rho = \rho_1$ and find
\be {dJ\o {d\rho}}(\rho_1)=  {{dJ\o {db}}(v_1)\o {d\rho\o {db}}(v_1)} =
{{1\o {v_1^2}} - \Big< {v\o {v-v_1}}\Big> \o {1\o {v_1^2}} \Big<{v\o{v-v_1}}
\Big>},
\ee
Thus here also the value of the parameter $l$ determines
the answer to the above question.\\
We should stress that the above arguments due to their qualitative nature
Is not by no means a substitute for exact solutions. However it is remarkable that in
view of the crude approximations involved, they can predict exact results.\\

\se{Acknowledgement} The authors wish to thank the hospitality provided
by the Abdus Salam International Center for Theoretical
Physics where this work was completed.
\se{Appendix}
For our presentation to be consistent, we have to show how the infinite dimensional
algebra (\rf{v4}-\rf{v4'''}) is derived in the MPA formalism.   
This can be simply done by a slight modification of the relations in \cite{k1}.
In the general case the Hilbert
space of each site of the lattice which we denote by ${\cal \bf h}$ is generated
by a discrete state $|0)$ (when the site is empty) and a continuous set of
states $|v)$, $v\in (0,\infty)$ (when the site is occupied by a particle of intrinsic
velocity $v$). We denote these states by different symbols to avoid confusion with
the states of the representations of the algebra. The states are normalized as
:
\be 
(0|0)=1\hskip 2cm (0|v)=(v|0)=0\hskip 2cm (v|v')=\delta(v-v')
\ee
The Hamiltonian is
\be
H = h_1 + \sum_{k=1}^{k=N-1} h^{B}_{k,k+1} + h_N.
\ee
where $h^B$ is given by 
\bea
h^B &=& -\int v \Bigg(|0 v)(v 0| - |v 0)(v 0| \Bigg)dv \cr
&-& \int \int_{v'>v}(v'-v)\Bigg( |v v')(v' v| -|v' v)(v' v| \Bigg) dv dv'
\eea
The boundary Hamiltonians $h_1$ and $h_N$ are:
\be
h_1 = -\int \a \sigma(v) v \Bigg(|v)(0| - |0)(0| \Bigg)dv 
\ee
\be
h_N = -\int (v+\b-1) \Bigg(|0)(v| - |v)(v| \Bigg)dv 
\ee
Note that the distribution $\sigma(v)$ only enters $h_1$, which 
points to the fact that the distribution $\sigma(v)$ refers to the 
particles injected to the system.

Inserting these Hamiltonians in the standard formulas of the MPA, i.e:
\be h^B {\cal A} \otimes {\cal A} = {\cal X} \otimes {\cal A} - 
{\cal A}\otimes{\cal X} \hskip 2cm <W| h_1 {\cal A} + {\cal X} = 0 
\hskip 2cm  h_N {\cal A} - {\cal X}|V> = 0
\ee
with the following form of the auxiliary vectors ${\cal A}$ and ${\cal X}$
\bea
{\cal A} &=& E |0) + \int \sigma(v) D(v) |v) dv \cr
{\cal X} &=& - |0) + \int \sigma(v) v |v) dv
\eea
leads to the algebraic relations (\rf{v4}-\rf{v4'''}).
Note that ${\cal A}$ and ${\cal X}$ are operator valued vectors in the Hilbert
space of one site of the lattice, as they should be in the MPA formalism.
\newpage
{\large {\bf References}}

\end{document}